\documentclass[%
aps,
pre,%
amsmath,amssymb,
preprint,%
]{revtex4-1}

\usepackage{graphicx}% Include figure files
\usepackage{dcolumn}% Align table columns on decimal point
\usepackage{bm}% bold math
\usepackage{color,soul}
\usepackage{subcaption}
\usepackage{amsmath}
\usepackage{comment}
\def \x{{\bf x}}

\def \rs{{\bf r}_s}
\def\x_s{{ \bf x_s}}

\def \f{{\bf f}}

\usepackage{graphicx}  % needed for figures
\usepackage{bm}        % for math
\usepackage{verbatim}   % for math
\usepackage{booktabs}
\usepackage{hyperref}

\begin{document}
	
%\preprint{AIP/123-QED}

\title{Sub-critical asymmetric Rayleigh breakup of a charged drop induced by finite amplitude perturbations in a quadrupole trap}
\author{Mohit Singh}
%\altaffiliation[]{ Department of Chemical Engineering, Indian  Institute of  Technology Bombay, Mumbai-400076.}
\author{Neha Gawande}
%\altaffiliation[]{ Department of Chemical Engineering, Indian  Institute of  Technology Bombay, Mumbai-400076.}
\author{Y. S. Mayya}%
%\altaffiliation[]{ Department of Chemical Engineering, Indian  Institute of  Technology Bombay, Mumbai-400076.}
\author{Rochish Thaokar}%
\altaffiliation[]{ Department of Chemical Engineering, Indian  Institute of  Technology Bombay, Mumbai-400076.}
%\homepage{https://rochishthaokar.wixsite.com/mysite}
\email{rochish@che.iitb.ac.in }
\date{\today}

\begin{abstract}
	The breakup pathway of Rayleigh fission of a charged drop is unequivocally demonstrated by first of its kind, continuous, high-speed imaging of a drop levitated in an AC quadrupole trap. The experimental observations consistently exhibited asymmetric, sub-critical Rayleigh breakup with an upward (i.e. opposite to the direction of gravity) ejection of a jet from the levitated drop. These experiments supported by numerical calculations show that the gravity induced downward shift of the equilibrium position of the drop in the trap cause significant, large amplitude shape oscillations superimposed over the center-of-mass oscillations.  The shape oscillations result in sufficient deformations to act as triggers for the onset of instability well below the Rayleigh limit (a subcritical instability). At the same time, the center-of-mass oscillations which are out of phase with the applied voltage, lead to an asymmetric breakup such that the Rayleigh fission occurs upwards via the ejection of a jet at the pole of the deformed drop. As an important application, it follows from corollarial reasoning that the nanodrop generation in electrospray devices will occur, more as a rule rather than as an exception, via asymmetric, subcritical Rayleigh fission events of micro drops due to inherent directionality provided by the external electric fields.
\end{abstract}
\pacs{Valid PACS appear here}
\keywords{Rayleigh breakup, charged droplet, high-speed imaging}
\maketitle

\section{Introduction}
A charged drop of diameter $D_d$, undergoes Rayleigh instability when the total charge on the drop exceeds a critical value, $Q_R=8 \pi\sqrt{\epsilon_e \gamma (D_{d}/2)^{3}}$, where, $\epsilon_e$ is  the electrical permittivity of the medium and $\gamma $ is the surface tension \cite{rayleigh1882}. At this critical charge the repulsive Coulombic force just balances the restoring surface tension force of the droplet. The Rayleigh instability is believed to be responsible for the breakup of raindrops in thunderstorms \cite{mason1972bakerian}, the formation of sub-nanometer droplets in electrosprays and generation of ions in ion-mass spectrometry \cite{fenn1989}. Although the theoretical limit of the critical charge is known for more than over hundred years \cite{rayleigh1882}, the breakup pathway was explicitly demonstrated only around a decade ago by Duft et al., (2003)\cite{duft03} through systematic experiments on a levitated charged drop in a quadrupolar trap. Their experiments indicate that a critically charged drop sequentially deforms to an elongated prolate spheroid eventually forming conical tips at its poles from which two jets are ejected out in the opposite directions. These jets carry 30-40\% of original charge and negligible mass ($\sim$ 1\%) \cite{duft03,giglio08}. The loss of charge reduces the electric stresses acting on the droplet and the deformed drop relaxes back to a spherical shape. The symmetrical jet ejection of a droplet tightly levitated in a quadrupole trap, may not correspond to practical situations such as electrosprays, wherein unbalanced external forces such as gravity or external electric field are most likely to introduce asymmetric breakup. The broken symmetry can have a tangible impact on the pathway of drop deformation as well as on the characteristics of daughter droplets formation. A significant body of theoretical literature, both analytical \cite{adornato1983shape,pelekasis1990equilibrium,natarajan1987role,tsamopoulos1985dynamics} and numerical\cite{basaran89,betelu06,das15,pelekasis1995dynamics,burton11}, indicates that instability of a charged droplet is subcritical with respect to ``finite amplitude" prolate spheroidal perturbations.
In this study, we provide an experimental evidence for asymmetric, sub-critical breakup through a combination of controlled observations and numerical simulations on levitated drops placed in external electric fields.

\section{Experimental setup}
\subsection{Materials and method}
\begin{figure}
	\begin{center}
		\begin{subfigure}{0.8\linewidth}
			\includegraphics[width=1\textwidth]{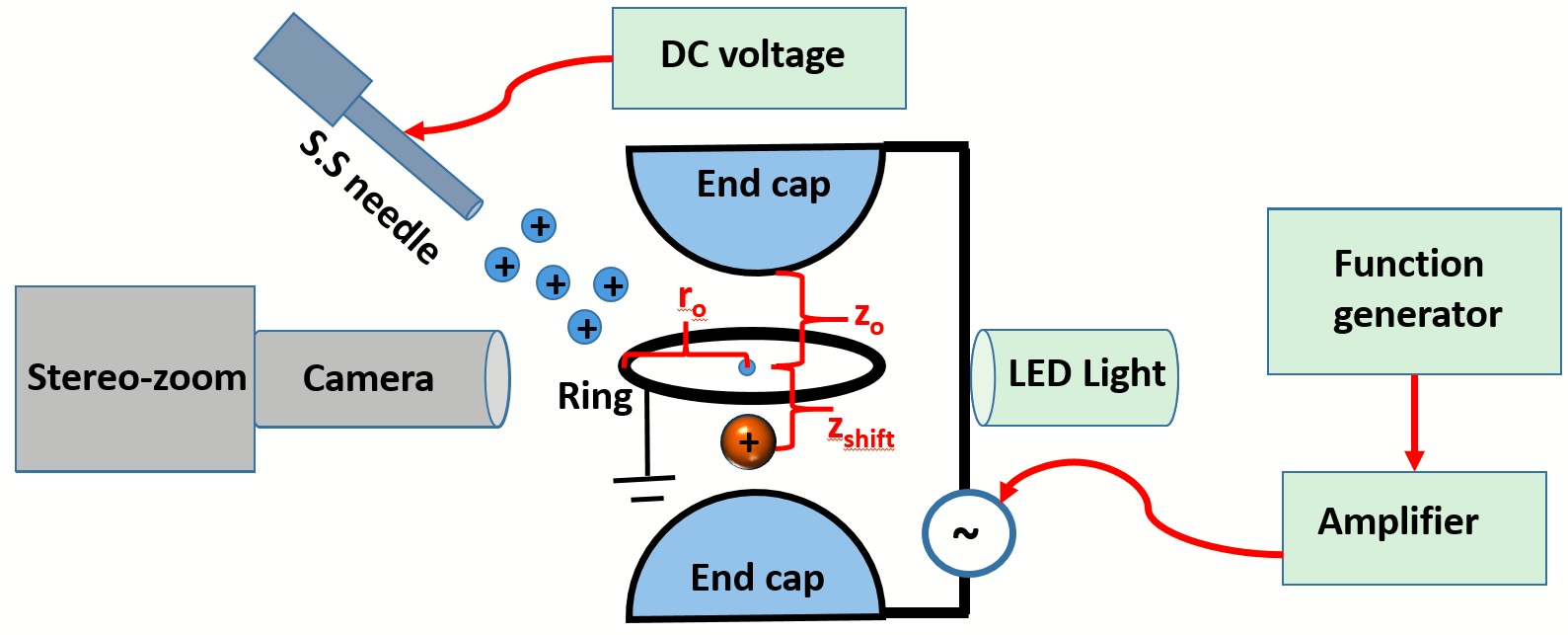}
			\caption{}
			\label{fig:setup}
		\end{subfigure}
		\begin{subfigure}{0.65\linewidth}
			\includegraphics[width=\textwidth]{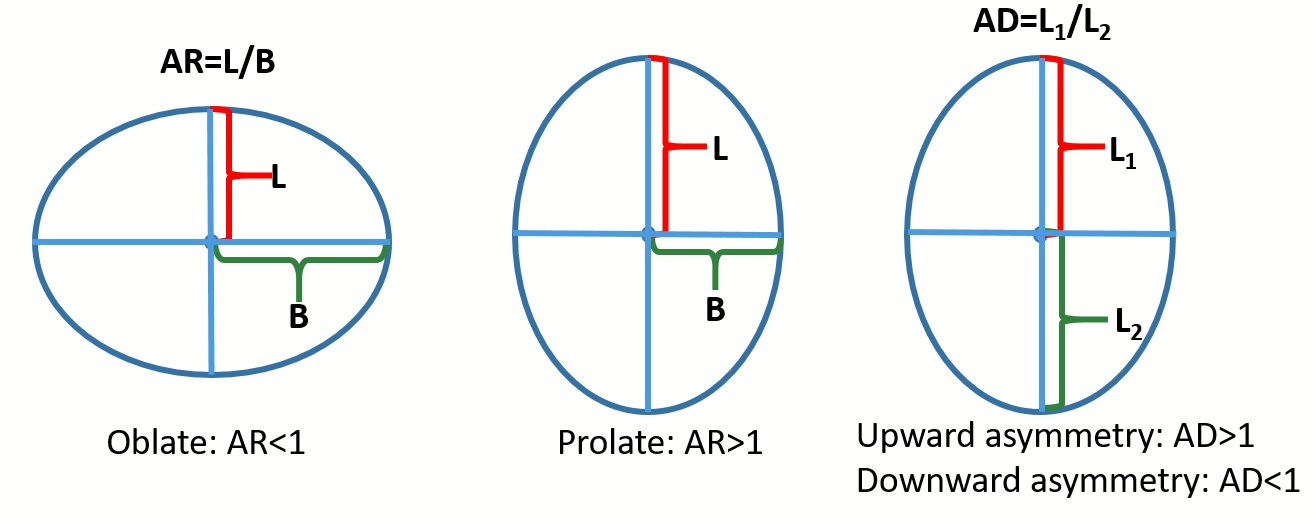} 
			\caption{}
			\label{fig:prolate_oblate}
		\end{subfigure}
		\caption{(a) Schematic of the experimental setup \textcolor{black}{used for studying} the generation, levitation and imaging of a charged drop in an electrodynamic balance, (b) Schematic of various lengths used for calculation of AR and AD.}
	\end{center} 
\end{figure}
The experiments were conducted by electrospraying (in dripping mode) a $positively$ charged droplet of an ethylene glycol-ethanol mixture (50\% v/v), into a quadrupole trap. \textcolor{black}{NaCl is added to increase the electrical conductivity ($\sigma$) of the droplet, which is measured using a conductivity meter (Hanna instruments, HI 2316), and the viscosity ($\mu_d$) of the droplets is measured using an Ostwald's viscometer. The surface tension ($\gamma$)} of the droplet is measured using a pendant drop (DIGIDROP, model DS) method, and the values obtained are reconfirmed with the spinning drop apparatus (dataphysics, SVT 20 ). The experiments are carried out at normal atmospheric conditions (1atm pressure and $25^o$ C temperatures is measured by using VARTECH instrument(THM-B2)). The temperature is maintained using air conditioner (capacity 1 ton, ECONAVI) installed in the experimental room (10$\times$10 ft). The relative humidity is measured using VARTECH instrument(THM-B2) and found to be around 20\%). 

The quadrupole trap used in the present experiments consists of two endcap electrodes, which are shorted and separated by 20 $mm$ ($=2 z_0$), and a ring electrode of the same diameter ($2 \rho_0$), as shown in figure \ref{fig:setup}. A function generator (33220A Function /Arbitrary Waveform Generator, 20 MHz) is connected to a high-voltage amplifier source (Trek, model 5/80, high-voltage power amplifier). This assembly is used to apply the potential of the desired waveform between the electrodes. The applied peak to peak AC potential in our experiments is 11$kV_{pp}$ with frequency varying from 0.1$kHz$ to 0.5$kHz$. 

In a typical experiment, charged droplets are generated using electrospray realized by applying high DC voltage (6-7kV) to a syringe tip. These charged droplets are then injected between the electrodes and are stabilized by the quadrupolar AC electric field between the endcap and the ring electrodes of the trap, resulting in electrodynamic levitation. Single charged drop levitation is achieved by simultaneously levitating a few drops in the trap by a series of injection and stabilization episodes. The process is rendered quite systematic by adjusting the potential applied to the syringe tip or by adjusting the trapping potential and frequency. After suspending the desired number of drops within the trap, the syringe tip potential is switched off, eliminating any further injection of the droplet cloud. A single droplet is made to survive in the trap, by eliminating (destabilizing) all the other drops by appropriately adjusting the driving frequency of the trap. The levitated single droplet is observed using a high-speed CMOS camera (Phantom V 12, Vision Research, USA), which is connected with a stereo zoom microscope (SMZ1000, Nikon Instruments Inc.). The camera is kept inclined at $30^0$-$40^0$ with respect to the plane of the ring electrode. The error in the droplet diameter due to camera inclination is observed to $\sim$2\%. Nikon halogen light (150W) is used as a light source.

\textcolor{black}{The shape deformations are characterized by two shape parameters; namely, aspect ratio (AR) and asymmetric deformation (AD), where AR indicates the symmetric deformation while AD is the measure of asymmetry in the shape of the drop. Thus, AR is defined as the ratio of the major axis (L) to the minor axis (B)} such that when AR$>1$, the shape is termed as prolate and when AR$<1$ the shape is called as oblate, as shown in figure \ref{fig:prolate_oblate}. While AD=$\L_1/L_2$, where, $L_1$ and $L_2$ are the distances of north-pole and south-pole from the centroid respectively (figure \ref{fig:prolate_oblate}). The charge on the droplet before and after the breakup is measured by the cut-off frequency method and also verified by the transient displacement method. The details of the methods can be found in our previous paper \cite{singh2017levitation}. \\

\subsection{Distribution of electric potential in quadrupole trap}
The potential of an ideal quadrupole trap is given by $\phi=\Lambda(z^2-0.5\rho^2)=\Lambda r^2 P_2(\cos{\theta})$, where $\Lambda=\Lambda_0\zeta(t)$, $\zeta(t)$ (e.g., $\cos(2\pi f t)$) is a time-periodic function of frequency $f$ while (z, $\rho$) and (r, $\theta$) stand for cylindrical and spherical polar coordinates respectively.  Here, $\Lambda_0=\phi_0/(\rho_0^2+2z_0^2)$ (where $\phi_0$ is the applied potential) is the intensity of an ideal quadrupole field. Unlike the case of the ideal Paul trap, where $\rho_0$=$\sqrt{2}z_0$, the present electrodynamic balance has $\rho_0$=$z_0$=10 mm. Since our electrodynamic balance is not an ideal Paul trap, the intensity of applied potential ($\Lambda_0$) is obtained by solving the electrostatic equation for the exact geometry of the setup in COMSOL Multiphysics software. The obtained data of the potential along $\rho$ and $z$ axes ($\phi(\rho,z)$) is then fitted into the equation of an ideal quadrupole trap by multi-linear regression method using a Origin (version 9.1.0 Sr2, b271) software. Thus the potential distribution in the $\rho$ and $z$ direction is obtained as:
\begin{equation}
\phi=(1.76\times10^7)[z^2-\frac{\rho^2}{2}]
\end{equation}
Thus the value of $\Lambda_0\sim 1.76 \times 10^7 V/m^2$ is used in the numerical calculations presented in next sections. 

\section{Model description}
To understand the mechanism of droplet breakup, numerical calculations are performed for a perfectly conducting liquid drop of radius $R$, suspended in a dielectric medium (air) in the presence of quadrupole electric field. In this study, Stokes equation for flow field and the Laplace equation for the electric potential ($\phi$) are solved using axisymmetric boundary integral method. Thus the governing non dimensional electrohydrodynamic equations can be written as,
\begin{equation}
\nabla \cdot {\bf{v}}_{d,a}=0,
\end{equation}
\begin{equation}
-\nabla {p_{d,a}}+ \chi_{d,a} \nabla^2 {\bf{v}}_{d,a}=0
\end{equation}
where, \textcolor{black}{subscript $d$ and $a$ represent drop (internal medium) and air (external medium) respectively.} ${p_{d,a}}$ is the  pressure, and $\chi$ denotes the viscosity parameter. Here, $\chi_a=1$ for external medium and $\chi_d=\lambda=\mu_d/\mu_a$ inside the drop.
\begin{equation} 
\nabla^2 \phi=0. 
\end{equation}
\textcolor{black}{Here, all the quantities in the units of length are non dimensionalised by $R_d$ (radius of drop) and the time, velocities and stresses are scaled by $\mu_d R_d/ \gamma$, $\gamma/\mu_d$ and $\gamma/R_d$ respectively, where $\gamma$ is the interfacial tension.} While the charge and electric fields are scaled by $\sqrt{R_d^3 \gamma \epsilon_a\epsilon_0}$ and $\sqrt{\gamma \epsilon_a\epsilon_0/R_d}$ respectively such that the non dimensional Rayleigh charge is $Q=8\pi$. \textcolor{black}{As the droplet conductivity is high ($\sigma$\textgreater $50 \mu S/cm$) in most of the} experiments, the ratio of charge relaxation timescale ($\epsilon_d \epsilon_0/\sigma_d$) to the hydrodynamic timescale ($\mu_d R/\gamma$) is quite small ($\sim$$\textless10^{-3}$). Hence, charge relaxation may be considered instantaneous and accordingly the droplet is modelled as a perfect conductor.

In this framework, the governing equations are then transformed into integral equations and are solved using standard methods reported in the previous papers \cite{deshmukh12,das15,gawande2017}. 
The integral equation of the electric potential for a perfect conductor drop is given by,
\begin{equation}
\phi({\bf r_s})=\phi_0({\bf r_s})+\frac{1}{4 \pi}\int \frac{E_{ne}({\bf r})}{|{\bf r-r_s}|} dS({\bf r})
\end{equation}
where ${\bf{r}}$ and ${\bf r_s}$ are the position vectors on the surface of the drop and $\phi_0$ is the applied electric potential which can be written as,
\begin{equation}
\phi_0(\rho,z)=\sqrt{Ca_\Lambda}[(z-z_{shift})^2-0.5\rho^2]
\end{equation} 
The unknown potential $\phi({\bf r_s})$ is constant on the surface of the drop, and is determined by the condition of conservation of charge given by $\int E_{ne} ({\bf r}) dS({\bf r})=Q$, where $Q$ is the constant surface charge on the drop. Since the breakup time is much smaller than the period of applied AC field ($\omega^{-1}$), the external potential (absorbed in $\sqrt{Ca_\Lambda}$) is assumed to be DC such that the end caps are at positive potential for a positive value of $\sqrt{Ca_\Lambda}$. The force density is then given by, $\Delta{\bf{f}}= {\bf{n}}\nabla \cdot {\bf{n}}-[\bf{\tau}_e]\cdot $, where $[{\bf{\tau}}_e]=(1/2) E_{ne}^2$ is the \textcolor{black}{normal} electric stress acting on the drop surface. A small shape deformation is introduced initially via a function of the form, $r_s(\theta)=(D_d/2)\Big{(}1+\sum_{l=1}^{4}\alpha_l P_l(\cos\theta)\Big{)}$, where $P_l$ is the $l^{th}$ Legendre mode and $\alpha_l$ is the corresponding coefficient. The force density is then used in the equation for integral equation of interfacial velocity which is given by,
\begin{equation}
{\bf{v}}(\rs)=-\frac{\lambda}{4\pi (1+\lambda)}\int \Delta\f({\bf r})\cdot G({\bf r},\rs)dS({\bf r})+\frac{(1-\lambda)}{4\pi(1+\lambda)}\int {\bf{n}}({\bf r})\cdot T({\bf r},\rs)\cdot {\bf{v}}({\bf r})dS({\bf r})
\label{eqn:velocity}
\end{equation}
where, $G(\bf r,r_s)=\frac{1}{|{\bf{x}}|}+\frac{{\bf{x} \bf{x}}} {{\bf{|x|}}^3}$ and $T(\bf r,r_s)=-6\frac{{\bf{x}}{\bf{x}}{\bf{x}}} {{|{\bf{x}}|}^5}$ are the kernel functions with ${\bf x}=(\bf r-r_s)$ and are extensively discussed in the literature \cite{sher88, pozri92}. The shape of the drop is then evolved with time using explicit Euler scheme. The details of the numerical scheme adopted in this study can be found in \citep{gawande2017}.  
\section{Results}
\begin{figure}[t]
	\centering
	\includegraphics[width=0.65\linewidth]{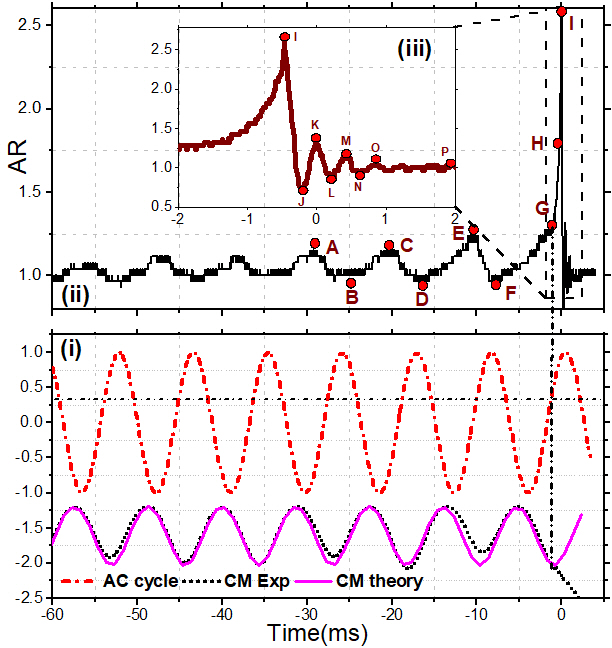}
	\caption{\textcolor{black}{The detailed mechanism of droplet centre of mass motion, surface oscillation, breakup and relaxation observed in a single high-speed video of a single drop, where, inset figures represent (i) experimental observation of the drop deformation dynamics in terms of AR, (ii) experimental and theoretical CM (obtained by solving modified Mathieu equation) oscillation dynamics in comparison with the normalized applied AC cycle and (iii) the enlarged region near the breakup and subsequent relaxation of the drop.} The intersection of verticle and horizontal black dash-dotted lines indicates the point of instability with respect to AC cycle. The red dots named from A-P are used to indicate the various stages of the droplet evolution and corresponding shapes are given in figure \ref{fig:sequence}. Parameters: $\mu_d$=6.0 mPa-s, $D_d=215\mu$m, $\gamma$=47 mN/m, $z_{shift}=500 \mu$m, $\mu_a=0.0185$ mPa-s, $f=114$ Hz, $\rho_d=960$ kg/$m^3$ and $\Lambda_0$=1.76$\times10^7$ V/$\text{m}^2$} 
	\label{fig:com_AR}
\end{figure}

\begin{figure}[t]
	\centering
	\includegraphics[width=0.8\linewidth]{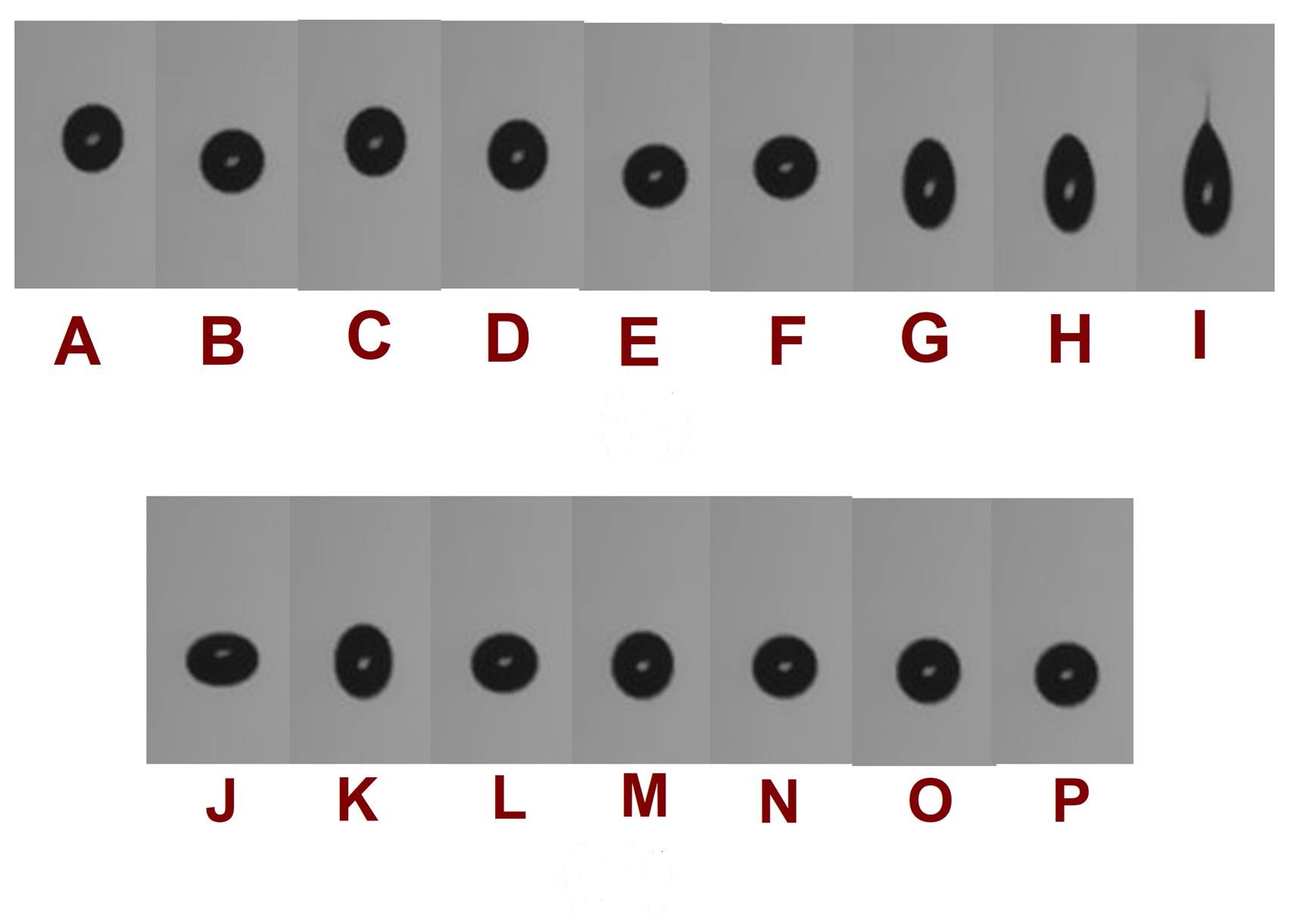}
	\caption{\textcolor{black}{Sequence of experimental images showing centre of mass motion, deformation, breakup and shape relaxation. The sequence of images A-F depicts CM and surface oscillations, images G-H show continuous deformation, image I indicates breakup and images J-P correspond to relaxation of the drop shape after breakup. The drop shape images are obtained by processing high-speed video in ImageJ software(\href{https://imagej.nih.gov/ij}{https://imagej.nih.gov/ij}). The parameters of the experimental observations are the same as given in figure \ref{fig:com_AR}.}} \label{fig:sequence} 
\end{figure}

In a typical experiment reported in this work, a charged droplet, under the influence of gravity, is levitated at an off-centered position in the quadrupole electrodynamic (ED) balance, as shown in figure \ref{fig:setup}. Unlike the ion trap in vacuum \cite{fenn1989}, which operates at GHz frequency, the present setup operates at sub kHz frequencies for levitating charged droplets at normal atmospheric pressure. It takes several minutes for the levitated droplets to evaporate to the point of attaining Rayleigh critical charge and undergo breakup. The events are recorded using a high-speed camera at speed in the range of 1500-2000 hundred frames per second (FPS) for about 2-4 seconds. The video is played back to analyse droplet center of mass (CM) oscillations (figure \ref{fig:com_AR}(i)), shape deformations and the asymmetric breakup event (figure \ref{fig:com_AR}(ii)) and its relaxation back to original spherical shape after the breakup (figure \ref{fig:com_AR}(iii)). \textcolor{black}{The image sequence shown figure \ref{fig:sequence} clear indicates all these stages of breakup.}

The droplet is seen to undergo simultaneous CM motion and shape deformations leading to an asymmetric breakup predominantly in the upward direction (that is at the north pole if the gravity acts from north to south). Out of the 49 breakup events observed, 42 cases resulted in upward ejection, whereas the rest exhibit downward ejection. About 24-30 \% charge loss is observed in the breakup, similar to that reported in the literature \cite{schweizer71,abbas67,roulleau72,doyle1964behavior,duft03,hunter2009}. The fact that a large number of frames captured in different stages of the charged droplet breakup process, observed through high-speed imaging of a single drop, makes it possible to compare the observations with continuous-time evolution models of the entire process. This constitutes a major distinguishing feature of this work. 

The video images \textcolor{black}{(see figure~\ref{fig:sequence})} raise four major questions: (i) How are the CM motion and the shape deformations related, and how do they affect the breakup pathway of the drop? (ii) Why does the droplet breakup predominantly \textcolor{black}{occurs} in one direction (upward, at the north pole)? (iii) How is the critical charge required to induce the instability modified due to CM-surface oscillations coupling? (iv) What is the role of the external quadrupolar potential on the droplet destabilization? To answer these questions, it is necessary to analyze all the stages observed in the breakup process, including oscillation mechanics of the drops in the quadrupole trap and is discussed in detail in the following sections.

\section{Discussion}
\subsection{Centre of mass motion}
The droplet in our experiments is levitated in a purely AC quadrupole field \cite{singh2017levitation}, unlike the previous study where the weight of the charged drop is balanced by an additional DC bias voltage \cite{duft02}. In the theoretical description of the problem, the weight of the droplet therefore appears in the $z$-directional (the direction of gravity) equation of motion, which is a modified Mathieu equation on account of the gravity and the frictional drag, as,
\begin{equation}
z''(\tau)+c z'(\tau)-a_z z(\tau) \cos(\tau)+\frac{g}{\omega ^2}=0,\label{matheiu}
\end{equation}
where, $a_z$=$2 \text{Q} \Lambda_0 (\tau)/(\frac{\pi}{6}D_d^3 \rho_d  \omega ^2)$, $c$=$3 \pi  D_d \text{$\mu_a$} /(\frac{\pi}{6}D_d^3 \rho_d \omega)$, $\tau(=\omega t)$ is the non-dimensional time, $\omega$=$2\pi f$, $\rho_d$ is the density of the drop, $\mu_a$ is the viscosity of the air, $Q$ is the charge on the drop. All the required parameters can be obtained from experiments, except the charge on the drop. 
With respect to CM stability of the droplet, two situations can arise. Firstly, a droplet can get loosely levitated with lower CM stability ($a_z$$\sim$0.25). With time the mass of the droplet continuously reduces due to evaporation, thereby increasing the value of stability parameter $a_z$. When $a_z$ reaches a critical value ($a_{z,critical}$$\sim$ 0.445, at $c$$\sim$ 0), the CM oscillations become violent (known as spring oscillations \cite{ataman1969measurement}) and the droplet tries to escape the ED balance. Thus to re-stabilize the droplet, the applied frequency is increased, which reduces the value of $a_z$. In most of the experiments (80\%), the frequency is adjusted such that the droplet is levitated at a value just below its critical stability limit ($a_z\sim 0.4-0.44$). Thus an approximate value of the charge on the droplet can easily be obtained from the definition of $a_z$. \textcolor{black}{The value of $Q$ calculated from the definition of $a_z$} using all other measured experimental parameters yields $Q$ of the order of Rayleigh charge ($Q_R$), clearly indicating that the droplet is charged near the Rayleigh limit.

\begin{figure}[t]
	\centering
	\begin{subfigure}[b]{0.45\linewidth}
		\includegraphics[width=\linewidth]{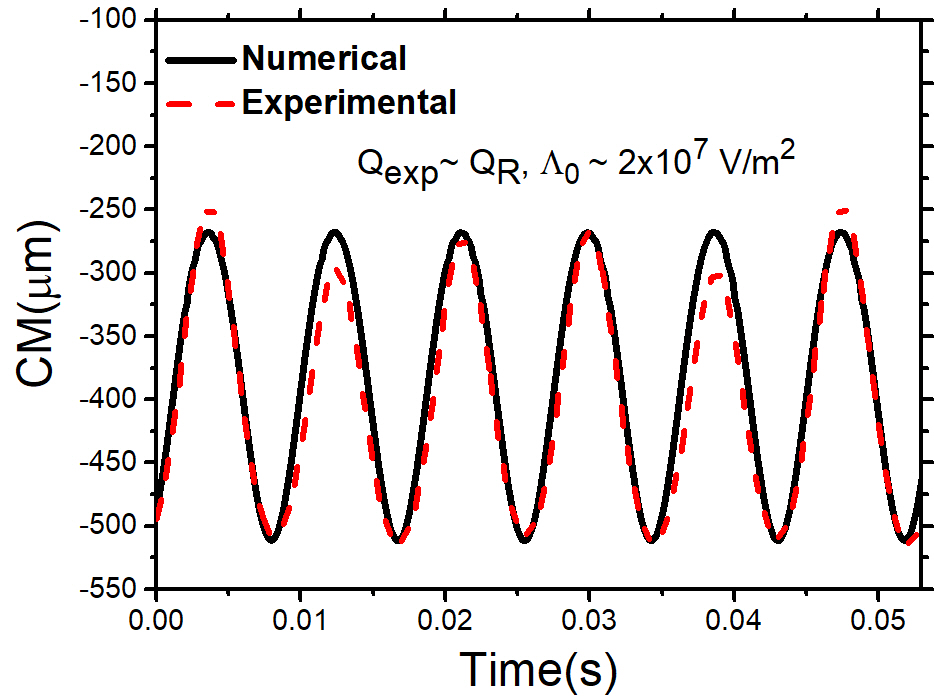}
		\caption{}
			\label{fig:com1} 
	\end{subfigure}
	\begin{subfigure}[b]{0.45\linewidth}
		\includegraphics[width=\linewidth]{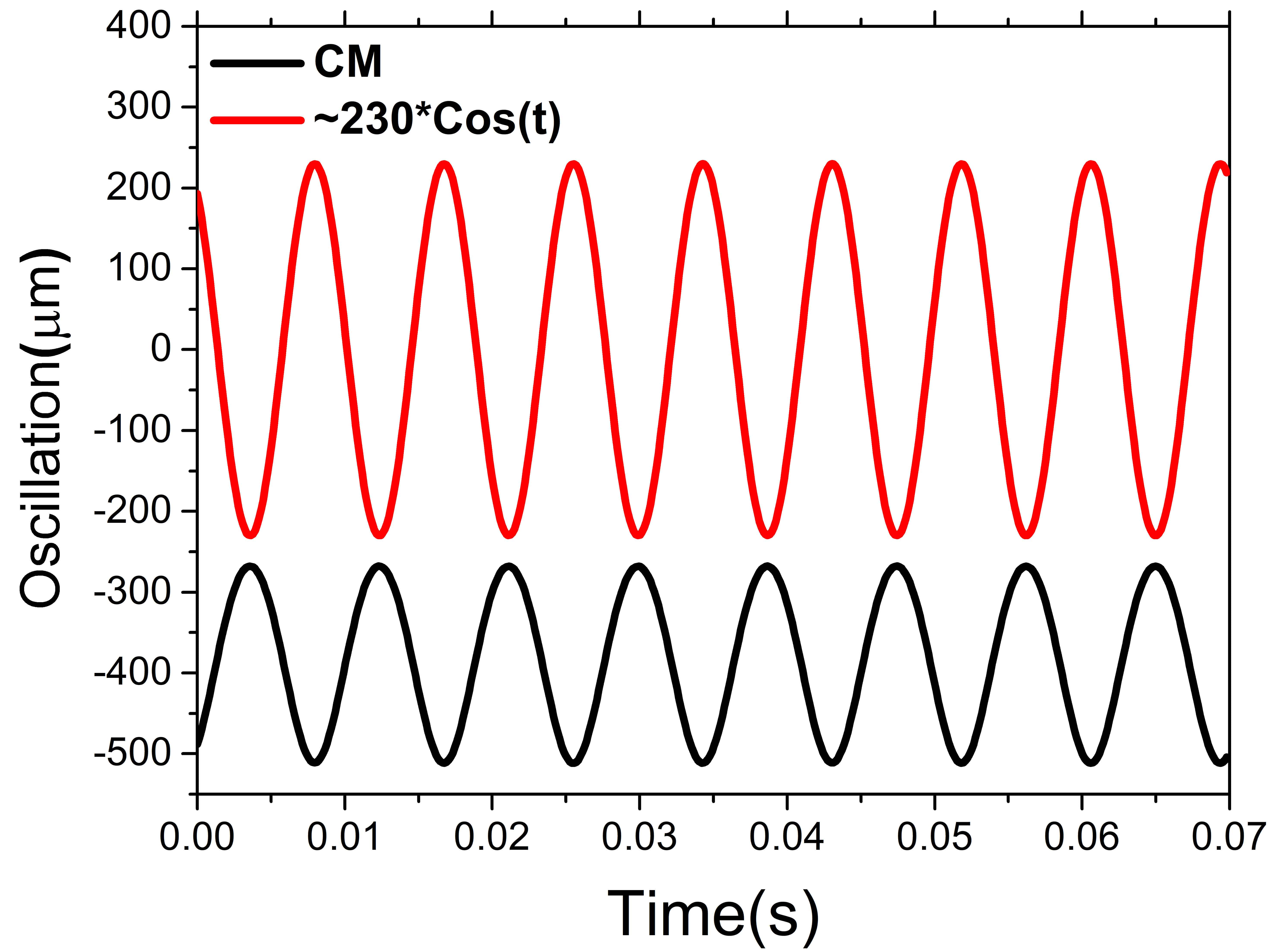}
		\caption{}
	\label{fig:com} 
	\end{subfigure}
	\caption{(a) Comparison of CM motion obtained from numerical solution of Eq.\ref{matheiu} and experimental image processing. \textcolor{black}{The values of parameters used for numerical solution are borrowed from experimental observation as indicated in the figure, (b) The variation of CM and applied AC field with time as obtained from the numerical solution.} The parameters used are the same as given in figure \ref{fig:com_AR}. } 
\end{figure}

In another situation, when a bigger sized droplet is levitated at a lower value of $a_z$, due to its high initial charge, the droplet breaks before it undergoes spring oscillations. In this case, the charge on the droplet can be estimated by fitting the value of the charge in the modified Mathieu equation to match the experimentally obtained amplitude of the CM oscillations, as shown in figure \ref{fig:com1}. It is interesting to note that the value of charge fitted to match the amplitude of the CM oscillations in the experiments is nearly equal to the Rayleigh limit of charge within $\pm$10\%  experimental error ($Q\sim Q_R$).

The CM oscillations can also be used to get the approximate $z$-directional shift ($z_{shift}$) of the drop from the center of the trap. From figure \ref{fig:com1} the maximum value of $z_{shift}$ estimated as $\sim$ 500$\mu$m. Since the droplet is found to oscillate between the $south$ $endcap$ and the center of the trap, the expression of a time-averaged equilibrium position at a distance $z_{shift}$ in terms of the trapping parameters can be obtained from the simple force balance in the $z$-direction as, 
\begin{equation}
m \omega^2\Big{(}\frac{1}{2}\frac{a_z^2}{1+c^2}z_{shift}\Big{)}=m g_z \label{ponderomotive}
\end{equation}
leading to $z_{shift}=\frac{2(1+c^2)g_z}{a_z^2\omega^2}$ \textcolor{black}{where, $a_z$ $\sim$ $2a_\rho$, $a_\rho$ is the stability parameter in the $\rho$-direction}. The left-hand side of Eq. \ref{ponderomotive} is the ponderomotive force which acts on the droplet towards the center of the trap (the details can be found in ref. \cite{singh2018theoretical}). The right-hand side of the Eq. \ref{ponderomotive} is the gravitational force ($mg$) acting on the droplet where $g_z$ is the acceleration due to gravity in the $z$-direction. This quasi-equilibrium, the time-averaged position, is a result of the balance between the ponderomotive force \cite{singh2018theoretical} and the gravitational force. The important observation here is that when the droplet is stabilized at an off-centered position, it oscillates with the applied frequency ($\omega$) around its equilibrium position (figure \ref{fig:com1}) experiencing a local uniform electric field ($E=2\Lambda Z_{shift}$) along with a non-zero quadrupolar electric field ($\Lambda$). The amplitude of these CM oscillations and $E$ are proportional to $z_{shift}$. In contrast, when the droplet is stabilized exactly at the center of the trap by annulling the force of gravity with DC field \cite{duft03}, it experiences negligible CM oscillations as well as the external influence of the quadrupolar fields. This apparently minor difference has a significant effect on the nature of droplet surface destabilization.    

The electrostatic force acting on a positively charged drop that exhibits oscillations in its CM motion critically depends upon the relative position of the charged drop within the trap, with respect to the oscillatory potential of the south end cap.
The numerical solution of Eq. \ref{matheiu} is plotted as a function of time along with the applied AC cycle, as shown in figure \ref{fig:com}. It can be observed that in the positive AC cycle, the position of the drop is lowest, i.e., near the south end-cap electrode (maximum negative displacement, also see figure \ref{fig:com_AR}(i)). This indicates that there is a phase shift of $\pi$ between the CM motion and the applied field. It will be seen later that the $\pi$ phase shift and the large amplitude CM oscillations of the droplet have an important implication on the asymmetric breakup of the droplet.
The exact phase lag between the applied AC field and the CM motion could not be measured in the experiments. The comparison is made possible by equating the peak position of the deformation (figure\ref{fig:com_AR}(ii)) to the negative peak of the AC cycle (figure\ref{fig:com_AR}(i)) at early times.
\begin{figure}[t]
	\centering
	\includegraphics[width=0.6\textwidth]{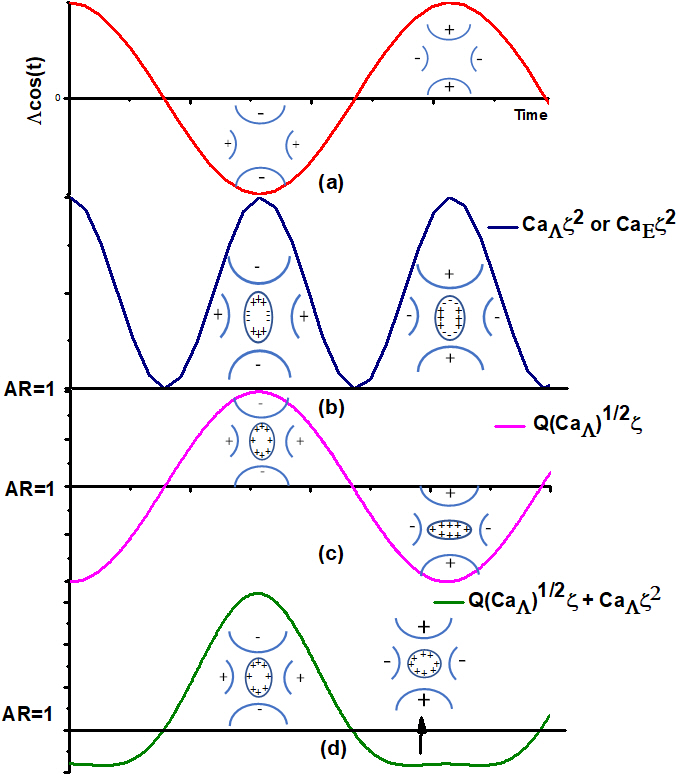}
	\caption{\textcolor{black}{A schematic representation of the effect of various electrical parameters on shape oscillation characteristics.}}
	\label{fig:sketch}
\end{figure}

\subsection{Surface oscillation dynamics}
The equations governing the surface dynamics, as well as the CM motion of the levitated droplet, are obtained from a leading order asymptotic theory for a charged droplet\cite{singh2018surface} in the quadrupole field. The surface of the droplet is described as, $
r_s(t,\theta)$ in terms of Legendre modes with $\alpha_l$ as the perturbation coefficient. The coefficient $\alpha_1$ represents the CM motion, $\alpha_2$, and $\alpha_4$ indicate symmetric dipolar and quadrupolar shape deformations, respectively, and the coefficient $\alpha_3$ is a measure of asymmetric shape deformation. The experimental observations of the oscillatory shape deformations can be described by a potential theory with viscous corrections \cite{singh2018surface} and these viscous corrections (which suppress high-frequency capillary oscillations) are found to be critical to explain the experimental observations.
$\alpha_1$ is given by Eq. \ref{matheiu} for CM, and the remaining  coefficients satisfy the following differential equations, to a linear order:

\begin{multline} \label{eqn:diff1}
\alpha_2''(t)+\frac{6}{3 \beta +2} \Bigr{(}2 (\lambda +4) \text{Oh} \thinspace \alpha_2'(t)+4 \alpha_2(t)- 3 \text{Ca$_E$}\thinspace \zeta^2+10 X \sqrt{\text{Ca}_\Lambda}\thinspace \zeta-\frac{25 \text{Ca}_\Lambda\thinspace \zeta^2}{7}\Bigr{)}=0 
\end{multline}
\begin{multline} \label{eqn:diff2}
\alpha_3''(t)+\frac{1}{4 \beta +3}\Bigr{(}24 (2 \lambda +5) \text{Oh}\thinspace \alpha_3'(t)+120 \alpha_3(t)-108 \sqrt{\text{Ca}_E} \sqrt{\text{Ca}_\Lambda}\thinspace \zeta^2\Bigr{)}=0 
\end{multline}
\begin{multline} \label{eqn:diff3}
 \alpha_4''(t)+\frac{1}{5 \beta +4}\Bigr{(}840 (\lambda +2) \text{Oh} \thinspace\alpha_4'(t)+2520 \alpha_4(t)-900 \text{Ca}_\Lambda\thinspace \zeta^2\Bigr{)}=0 
\end{multline}
	
where, \textcolor{black}{$Oh=\mu_d/\sqrt{\gamma (D_d/2) \rho_d}$} is the Ohnesorge number, $\beta=\rho_a/\rho_d$ is the density ratio between the droplet and the surrounding medium, $X$=$Q/Q_R$ is the fissility, $Ca_\Lambda=(D_d^3 \epsilon_e/8 \gamma)\Lambda_0^2$ and $Ca_E=(D_d \epsilon_e/2 \gamma)E^2$, where $Ca_\Lambda$, $Ca_E$ \textcolor{black}{are the force due to quadrupolar potential and uniform fields respectively}. The effect of several electrical force terms such as $\sqrt{Ca_E} \sqrt{Ca_\Lambda}$ (asymmetric force on an uncharged drop due to uniform and quadrupole field coupling), $Q \sqrt{Ca_\Lambda}$ (force on a charged drop due to quadrupole field) and $Q\sqrt{Ca_E}$ (force acting on a charged drop due to uniform field) on the characteristics of surface dynamics (figure\ref{fig:com_AR}(i)) can be understood by solving Eq. \ref{eqn:diff1}, \ref{eqn:diff2} and \ref{eqn:diff3} simultaneously. A schematic representation of effect of various parameters/forces on the characteristics of a charged droplet oscillations is shown in figure \ref{fig:sketch}. The figure \ref{fig:sketch}(a) is a variation of the  applied potential with respect to time. If the value of $\Lambda(t)$ is positive the end-cap potential is positive and vice-versa.

The shape oscillations of an off-centered $uncharged \thinspace drop$ in a quadrupole trap are caused by Ca$_{\Lambda}$ and Ca$_{E}$. Ca$_{\Lambda}$ induces both dipolar and quadrupolar shape oscillations with a frequency of $2 \omega$ while, Ca$_{E}$ (due to off-center position) excites dipolar oscillations with a frequency $2 \omega$. These forces induce the polarization of the free charges in a neutral drop, and the droplet oscillates with the frequency of $2 \omega$ (figure \ref{fig:sketch} (b)). \textcolor{black}{ On the other hand, the quadrupolar field and the uniform field, independently acting on the total unperturbed charge ($Q$) of the undeformed drop lead to symmetric shape deformations with frequency $\omega$ due to $\sqrt{Ca_{\Lambda}} Q$ term  as shown in figure (\ref{fig:sketch}(c)) whereas $\sqrt{Ca_{E}} Q$ causes the translation of the drop.}. The experiments (figure \ref{fig:com_AR}(i)) indicate that the shape oscillations occur at frequency $\omega$ with weak oblate deformations thereby suggesting a complex interplay between the terms  $\sqrt{Ca_{\Lambda}} Q$, Ca$_{\Lambda}$ and Ca$_{E}$. It may be inferred from the equation \ref{eqn:diff1} that, if the relative magnitude of $\sqrt{Ca_{\Lambda}} Q$ dominates over other parameters, then the droplet surface oscillates symmetrically with the applied frequency ($\omega$). On the other hand, if the magnitude of $Ca_{E}$ and Ca$_{\Lambda}$ dominate, the resultant shape oscillation (schematic figure \ref{fig:sketch} d) is similar to the experimental observation shown in figure \ref{fig:com_AR}(ii) (oscillations from -60ms to -30ms). The shape deformations can also be understood in terms of the electrostatic attraction or repulsion between the charge on the droplet and the polarity of the electrodes. When the applied potential is positive (and maximum), the positively charged droplet experiences maximum electrostatic repulsion from the end cap electrodes and deforms the droplet into an oblate shape ($AR\textless1$). Similarly, in the negative AC cycle, the droplet experiences electrostatic attraction from the end cap electrodes and deforms into a prolate spheroid ($AR\textgreater1$)(figure \ref{fig:sketch} (c)). Using this reasoning, the peak negative potential of the AC cycle is made to coincide with the peak prolate amplitude of the shape deformation observed in the experiments, so as to match the time history of deformation and the applied field. \textcolor{black}{The asympototic analysis also indicate that the oscillation pattern obtained due to the complex interplay between the terms $\sqrt{Ca_{\Lambda}} Q$, Ca$_{\Lambda}$ and Ca$_{E}$ can induce mode coupling in the drop shape which in turn can significantly affect the breakup mechanism of the droplet.} \\

\subsection{Droplet Breakup}
\begin{figure}[t]
	\centering
	\includegraphics[width=0.8\textwidth]{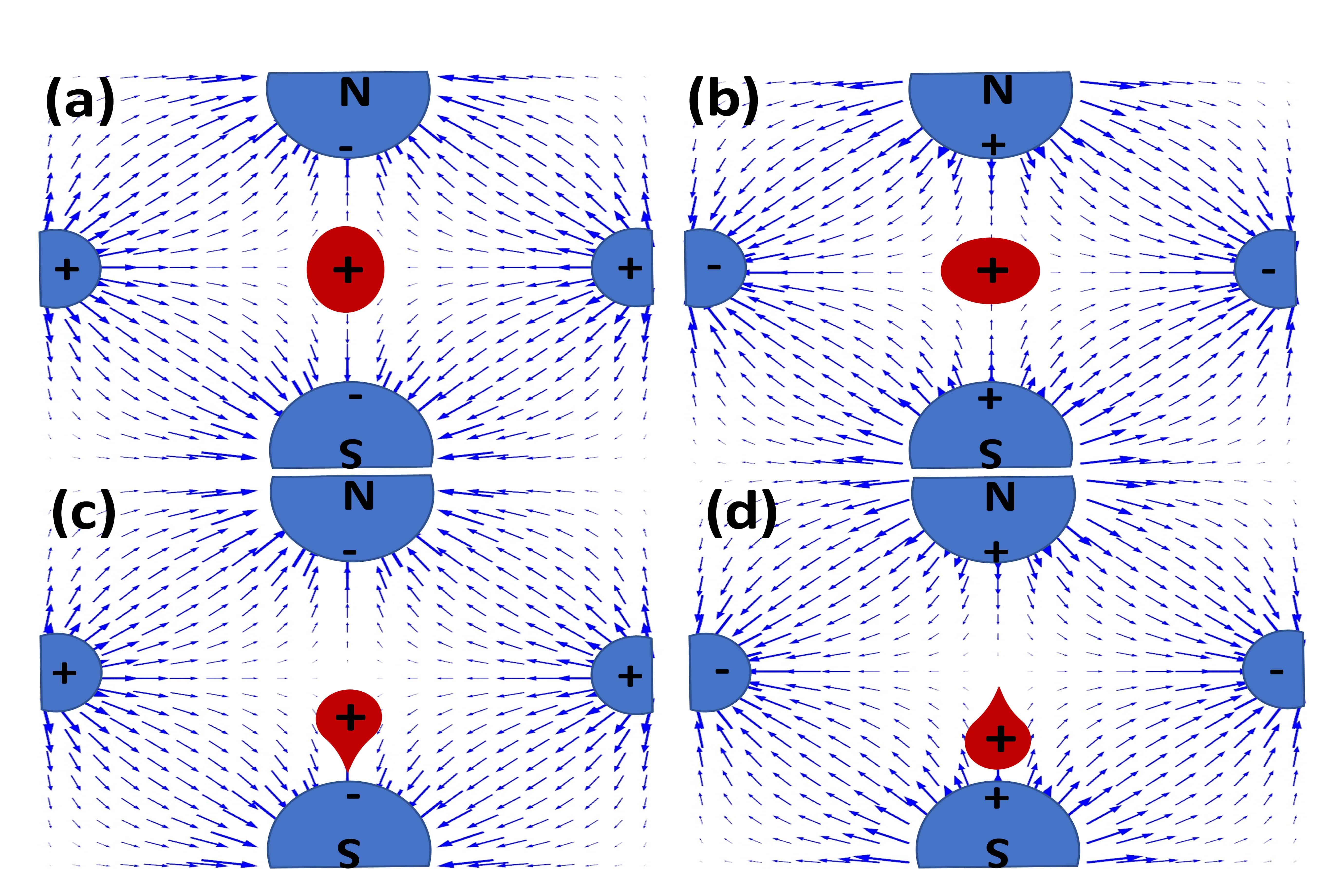}
	\caption{Schematic representation of polarity of electrodes, droplet position and corresponding deformation leading to breakup. Figures (a) and (b) show drop at centre of the trap and symmetric deformation, whereas figures (c) and (d) show drop at off-centred position with endcap polarity dependent deformation and breakup.}
	\label{fig:field}
\end{figure}

In the course of executing both the CM and shape oscillations, the evaporation process causes the droplet charge to approach its Rayleigh limit, leading to the onset of Rayleigh instability, which eventually causes an asymmetric breakup via the formation of a jet. The sizes of the ejected droplets could not be measured very accurately due to the limitation of the resolution of the microscopy used in our experiments. However, relook at the hazy droplet images provides an approximate estimate of the droplet size as $\sim$ 9 $\mu$m. The daughter droplets move away from the observational field of view within 10-20 $\mu$s after their ejection due to an electrostatic repulsion from the mother droplet.  As a result, it is not possible to make direct observations on the fate of the daughter droplets. However, based on the observations and analysis of the breakup pathway of the mother droplet, it is reasonable to expect that the ejected daughter droplets would undergo further break up giving rise to yet smaller satellite droplets, and so on.
Since an AC quadrupole field is used for levitation of a positively charged drop, the relative potential (either positive or negative) of the end cap and the corresponding deformation are critical to asymmetric breakup.
While the quadrupolar field $\Lambda_0$ corresponding to that acting at the center of the drop can only induce symmetric deformation in the \textcolor{black}{drop thereby a possible symmetric} breakup (as shown in figure \ref{fig:field}(a),(b)), any asymmetric breakup should occur due to the differential, locally uniform field $E$ acting on the surface of the droplet (as shown in figure \ref{fig:field}(c),(d)).
A positively charged droplet near a positive south end cap should deform into an oblate spheroid due to the electrostatic repulsion at the poles between the like-charged drop and the end cap as well as due to the electrostatic attraction between the oppositely charged drop and the ring electrode at the equator (figure \ref{fig:field}(d)). On the other hand, if a positively charged drop is near the negative south end cap (figure \ref{fig:field}(c)) it should break in the downward direction due to higher electrostatic attraction from the south end cap and repulsion from the ring electrode (as shown in figure \ref{fig:field}(c)). 

 The experiments show that the breakup is predominantly asymmetric, with a majority of jet ejection events occurring upwards (at the north pole, against gravity, as shown in figures \ref{fig:sequence} and \ref{fig:field}(d)). An analysis of the CM motion on the levitated charged droplets indicates that most often, the droplet breakup occurs when the droplet is near the \textcolor{black}{south }end-cap electrode (see Supplementary 2). Experiments  (figure \ref{fig:com_AR}(i)) also show that the positively charged droplet is in the vicinity of the south end cap when the latter is at positive peak potential. This is attributed to the phase lag of $\pi$ between the applied potential and the CM motion. The predominance of the upward breakup is somewhat intriguing, and its understanding requires careful theoretical analysis of the problem.
 
 To further understand the underlying mechanism and the stresses responsible for the asymmetric breakup, we performed numerical simulations using the axisymmetric boundary integral (BI) method. From the asymptotic analysis in small $\Lambda$, (Eq.\ref{eqn:diff1}, \ref{eqn:diff2}, \ref{eqn:diff3}), it is observed that the viscosity of the liquid drop plays an important role in the surface oscillation dynamics of a charged drop in terms of diminishing the effect of the natural frequencies associated with different modes. Also, at the onset of jet ejection, the rate of change of AR values predicted by the simulations (presented in our previous work \cite{gawande2017}) in the viscous limit is in good agreement with those observed in the previously reported experiments \cite{giglio08}. Thus, to understand the mechanism of droplet deformation and breakup, BI calculations are carried out in the viscous flow limit. Experiments indicate that the breakup of the droplet occurs in time that is $(1/4)^{th}$ the period of the AC cycle of the applied field. In view of this, the simulations are carried out by considering either positive or negative DC quadrupole potential with the intensity of the applied electric field $\Lambda_0$. All the parameters are borrowed from direct observations of the experiments, and the value of $z_{shift}$ is taken as the maximum displacement observed from figure \ref{fig:com1} i.e., $\sim$ 500 $\mu$m. 
	
\begin{figure}[t]
	\centering
	\includegraphics[width=0.8\textwidth]{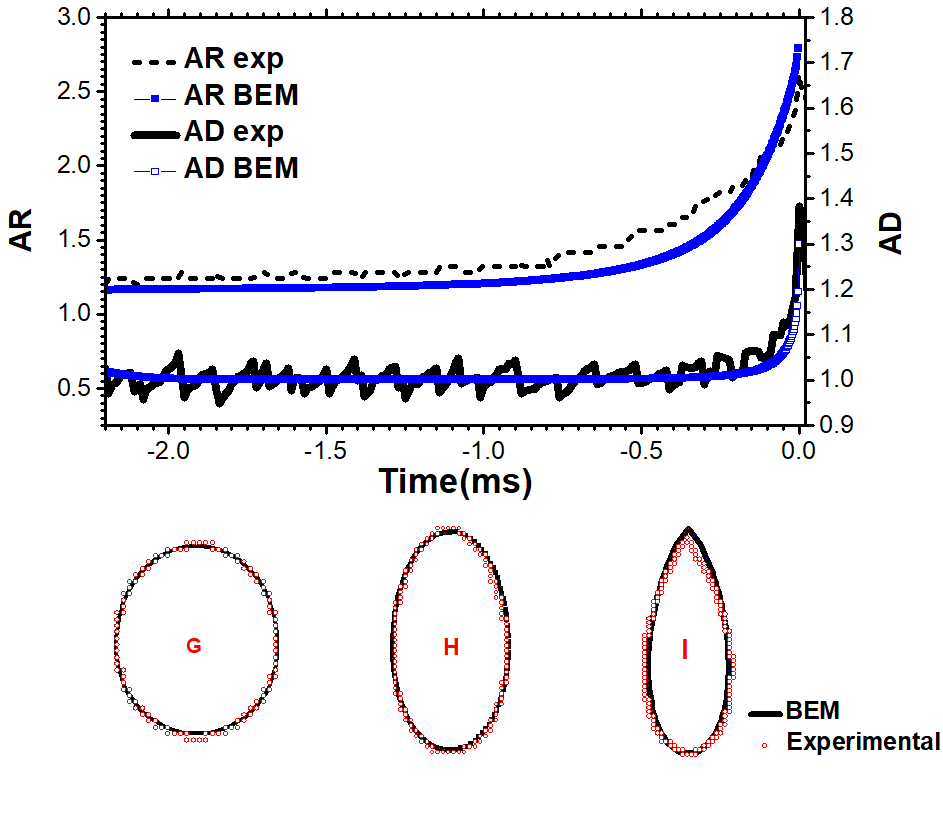}
	\caption{\textcolor{black}{Comparison of experimental observations with BI simulations for temporal evolution of AR and AD along with the drop shapes at three different times near the breakup (shown in the bottom panel).}}
	\label{fig:AR_AD_match}
\end{figure}	
We first conduct calculations with an initially spherical droplet. The simulations indicate that at Rayleigh charge (i.e., 8$\pi$) with a downward $z_{shift}$ and an unperturbed initial spherical shape, the drop breaks in the downward direction for a negative end-cap potential. The observation is explained by the fact that in the presence of negative end-cap potential, the positive charges get accumulated toward the south pole of the drop due to electrostatic attraction between charges on the drop and the end-cap electrode. Due to the accumulation of charges, the droplet develops high curvature at the south pole, which leads to a downward breakup for experimentally relevant parameters. However, in the case of a positive end-cap potential with downward $z_{shift}$, the droplet experiences an electrostatic repulsion from the end-cap electrode as well as attraction from the ring electrode. Thus, in this configuration, the droplet renders a stable oblate shape and cannot undergo breakup for typical experimental parameters.   

These results contradict the experimental observation where the droplet breaks in the upward direction in the positive cycle of the applied field (which corresponds to positive end-cap potential). The apparent inconsistency can be resolved by observing that in figures \ref{fig:com_AR}(ii) and \ref{fig:sequence}, at point G, which is the droplet state just before the breakup, the droplet exhibits a highly deformed prolate spheroidal shape and does not exhibit prolate to oblate oscillations thereafter. As at this point, the droplet has built a critical charge to admit Rayleigh instability that leads to a breakup. We, therefore, consider this point to be the onset of the Rayleigh instability, and the experimentally observed drop shape corresponding to image G in figure \ref{fig:sequence} is considered as an initial shape in the numerical calculations. The outline of the drop shape in image G is obtained using the ImageJ software and is fitted using the nonlinear least square method to a Legendre series (using Mathematica software, version 10) to obtain the coefficients of the different Legendre modes. The details of shape fitting analysis can be found in Supplementary 2. The coefficient of second Legendre ($P_2$) mode is thus obtained as 10.56 $\mu$m and that of third Legendre ($P_3$) mode as +2.08 $\mu$m and the radius ($D_d/2$) of undeformed drop is 108 $\mu$m. The shape fitting indicates that the symmetric $P_2$ mode is most prominent. Incidentally, $P_2$ is the most unstable mode as predicted by the linear and nonlinear analysis of Rayleigh breakup of a  charged drop \cite{basaran89, thaokar2010}. A significant value of the asymmetric $P_3$ mode is also observed (a positive value of $P_3$ mode means a higher curvature at the north-pole and vice-versa). The simulations are initialized with a shape corresponding to image G (figure \ref{fig:sequence}) where the initial shape of the droplet considered in the simulations is perturbed with the coefficients of $P_2$ and $P_3$ modes, obtained from the experiments. For numerical simulations the parameters are non-dimensionalized with $D_d/2$ and are given as: $Ca_\Lambda$=0.00058, $z_{shift}$=4.63, $\alpha_2$=0.1, $\alpha_3$=0.02.

\begin{figure}[t]
	\centering
	\includegraphics[width=0.8\textwidth]{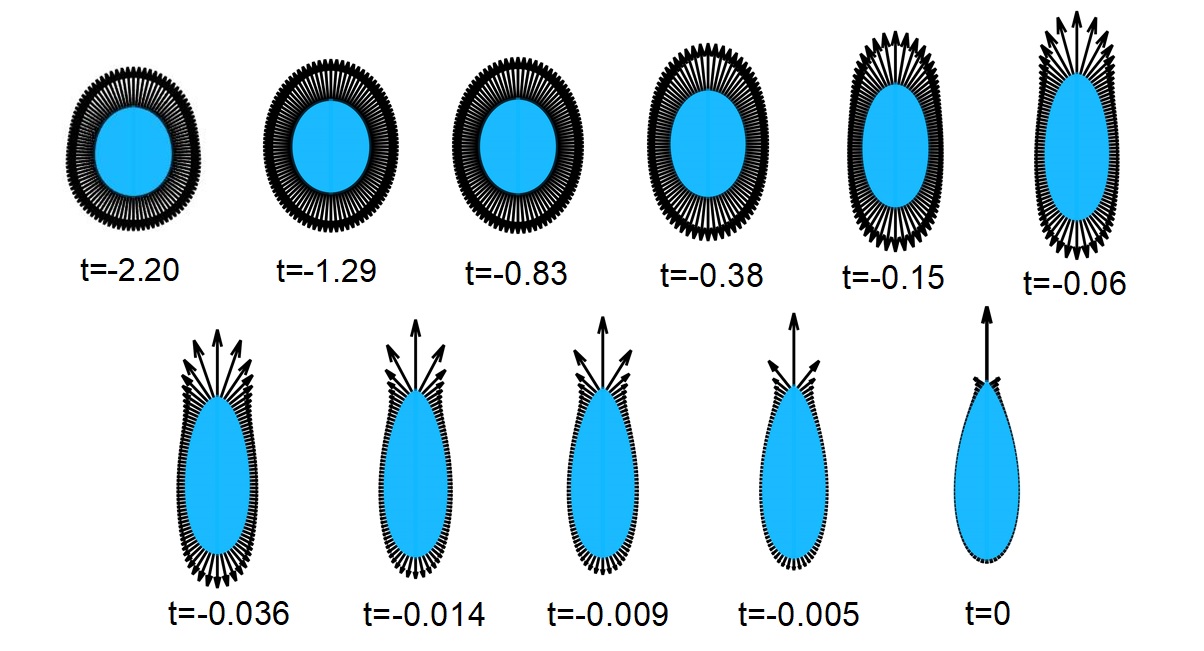}
	\caption{The stress distribution on the surface of the drop obtained from BI simulation. The numbers indicated below the drop shapes are times in millisecond.}
	\label{fig:stress}
\end{figure}

\textcolor{black}{The BI calculations are carried out by providing an initial surface charge $Q$ to the perturbed (the perturbation is obtained form the experimental drop shape) drop }at a value which is in the sub-Rayleigh charge limit and progressively increasing it till the critical (i.e., minimum) value of the charge at which the droplet undergoes breakup. If the charge is less than this critical value, the droplet relaxes back to the spherical shape. However, when the charge is beyond the critical value, the initial shape perturbations grow, and the droplet evolves to form sharp conical ends, finally admitting a numerical singularity\cite{gawande2017}. It is found that the droplet breaks at 98.7 \% (i.e., 7.9$\pi$) of the Rayleigh charge for the given parameters. The breakup at subcritical charge clearly demonstrates that subcritical Rayleigh instability can be induced by finite-amplitude perturbations.  Since the external electric field is small, it merely acts as a trigger for inducing surface perturbations and has \textcolor{black}{insignificant} role in causing break up.  This is in line with the prediction of the theory that the breakup of a droplet in the quadrupole electric field is a transcritical bifurcation \cite{das15}. 

A closer look at the role of initial perturbation suggests that for an initial shape with dominant $P_2$ perturbation and a positive $P_3$ perturbation, the breakup is always observed in the upward (at the north pole) direction when the end cap is positive, in conformity with experiments. Under these conditions, there can be two mechanisms responsible for asymmetry in the droplet shape at the onset of the breakup. Firstly,  a significant positive $\alpha_3$ perturbation (for the $P_3$ mode) in the initial droplet shape can assist upward and asymmetric breakup. 
Secondly, an $\alpha_3$ perturbation can develop due to the nonlinear interaction between the dipolar charge distribution on the drop and the positive (when the south end cap is positive) uniform electric field ($E$) experienced by the prolate spheroidal droplet. Mathematically, the $P_3$ term may be understood as arising out of the nonlinear coupling between $P_2$ and $P_1$ (i.e., dipolar charge distribution and uniform electric field $E$). The $P_3$ mode leads to an upward (north pole) jet formation and subsequent breakup. Since the magnitude of $\alpha_3$ depends on the strength of the nonlinear coupling between $\alpha_2$ and $E$ (which takes a certain amount of time to become significant), the asymmetry manifests at a later stage in the dynamics of the drop breakup.  This explains the late onset of asymmetric breakup seen in figure \ref{fig:AR_AD_match}. Although both these mechanisms may act simultaneously in assisting upward breakup, the latter is seen to be more dominant than the former.

The above reasoning may be quantified by examining the time evolution of normal stresses acting on the surface of the drop (see figure~\ref{fig:stress}) using the BI method. Initially, when the drop is perturbed with experimentally obtained values of shape deformation coefficients, the normal electric stresses acting on the drop surface are only marginally higher at the north-pole. As may be seen in figure \ref{fig:stress}, the stress distribution remains nearly symmetric for a considerable time (between t= -2.2 to -0.15 ms). The nearly symmetric stress distribution also corresponds to nearly symmetric dipolar ($P_2$) charge distribution. Beyond this point (t=-0.15), the coupling between $P_1$ and $P_2$ escalates, creating an asymmetric ($P_3$) charge distribution and thereby asymmetric Maxwell stress distribution, which through a feedback mechanism manifests as an asymmetric upward breakup. The asymmetry is thus a result of the finite-amplitude perturbation of the symmetric mode, and therefore the AD manifest later (as compared to AR) in the instability. The mechanism prevails even when the initial perturbation does not have the $P_3$ mode ($\alpha_3=0$), thereby explaining why the upward breakup is more prevalent in experiments. 
The droplet shape evolution predicted by BI simulations is compared with the experimental observations, as indicated by AR and AD. A remarkable agreement is observed between the experiments and the simulations (figure \ref{fig:AR_AD_match}), including the important observation of a late onset of asymmetry. Moreover, the shapes of the drop from the critical point (G) to the breakup point (I) are accurately predicted by the BI simulations, as shown in figure \ref{fig:AR_AD_match}. \textcolor{black}{The drop shape corresponding to the image I is compared without considering the jet part observed in the experiments since the BI calculations for perfect conductor drop cannot capture the jet formation. To predict the jet and progeny formation in the breakup of charged drops, it is necessary to consider the finite charge dynamics on the drop surface \cite{gawande2020jet}. Since in this study, BI simulations are carried out for the understanding of the underlying mechanism of the experimentally observed asymmetric breakup of highly conducting charged drops the finite surface charge dynamics is neglected.} 
In certain cases (in about 10\% of the cases) where $\alpha_3$ is negative, a downward breakup has been observed. This can be explained as follows: although a phase shift of $\pi$ is observed between the applied AC field and the CM oscillations, in few experiments, we observe that the south endcap is at negative potential when the droplet is at the bottom of its CM motion and attains a critical shape. Moreover, the shape at this stage also indicates a negative $\alpha_3$ (for details see Supplementary 2). Since this situation is rare, fewer droplets are found to break in the downward directions. 
 
\section{Conclusions}
 While the work of Duft et al. \cite{duft03} has shown pioneering evidence for a symmetrical pathway for Rayleigh breakup, the present study shows that this is not universal, and perhaps is an exception in realistic situations. Specifically, by considering the effect of gravity and external electric fields and from the insight obtained by numerical calculations, an asymmetric breakup might turn out to be the rule in real-life practical situations, such as the one that is commonly encountered in nano-drop generators using electrosprays.  This is amply demonstrated through continuous high-speed imaging of levitated drops in a quadrupole trap combined with BI simulations. The external electric fields act as initiators of finite-amplitude shape deformations, which assist in driving the droplets towards sub-critical Rayleigh breakup even when the charges carried by them are less than the  Rayleigh critical charge. BI calculations accurately predict these values to be about 98-99\% of the Rayleigh limit in conformity with experimental observations. It must be pointed out that detailed bifurcation diagrams exist for Rayleigh instability, and these clearly include transcritical bifurcations at $Q=Q_R$ \cite{basaran89, das15} as well as imperfect transcritical bifurcations in the presence of walls \cite{mhatre12}.  The prolate and oblate deformations correspond to sub and supercritical bifurcations, respectively.  The present observations are in conformity with a previous study that showed that the applied quadrupole field could further reduce the critical Rayleigh charge due to an interaction between the applied field and the Rayleigh instability \cite{das15}.  The external fields also induce asymmetric jet ejection, and the study shows that there is a strong shape instability coupling in the dynamics of charged drops. 
The study has a far-reaching bearing on technologies exploiting Rayleigh break up process for nanoparticle generation using electrosprays, or ion mass spectrometry. The occurrence of subcritical break up will influence the time history, and asymmetric break up will affect the evolution of the spatial distribution of droplet sizes. Moreover, judicious choice of the polarity of confining electrodes can lead to greater effectiveness in the breakup of droplets. All these practical implications of the findings presented here deserve careful considerations in future studies.

\section*{Author contributions statement}

Mohit Singh performed the experiments, data analysis  and writing of the manuscript.
Neha Gawande performed BI simulations, data analysis and writing of the manuscript.
Y. S. Mayya contributed to application of analytical tools to data analysis and review of the manuscript.
Rochish Thaokar conceived the problem, analyzed and interpreted the data and reviewed the manuscript. 

\section*{Additional information}
The additional information is provided in the separate supplementary files.
Supplementary 1 is the high-speed videographic evidence of sub critical asymmetric Rayleigh breakup event. Supplementary 2 contains details of shape fitting analysis, CM oscillations and few more experimental observations for upward and downward breakup. 


%\bibliography{refss2}

\end{document}